\documentstyle[11pt,psfig,emulateapj]{article}

\def\pcmsq{{$\rm cm^{-2}$}}
\def\fxu{{$\rm ergs\ cm^{-2}\ s^{-1}$}}

\def\rchi{{$\chi^{2}_{\nu}$}}
\def\delchi{{$\Delta \chi^{2}$}}

\def\nh{{$N_{\rm H}$}}

\newcommand{\lta}{\lesssim}
\newcommand{\bl}{{BL Lacs\ }}
\newcommand{\vel}{\rm km\ s$^{-1}$}

\def\lesssim{\, 
\lower2truept\hbox{${<\atop\hbox{\raise4truept\hbox{$\sim$}}}$}\,}
\def\gtrsim{\, 
\lower2truept\hbox{${> \atop\hbox{\raise4truept\hbox{$\sim$}}}$}\,}

\received{14 Oct 1999}
\accepted{28 Feb 2000}
\journalid{538}{July 20 2000}
\slugcomment{To appear in The Astrophysical Journal, July 2000}

\begin{document}

\title{ROSAT/ASCA Observations of a Serendipitous BL Lac Object PKS 2316-423:
       The Variable High-Energy Tail of Synchrotron Radiation}
%\altaffiltext
\author{Sui-Jian Xue\altaffilmark{1}, You-Hong Zhang\altaffilmark{2}, 
\and Jian-Sheng Chen\altaffilmark{1}}
\altaffiltext{1}
{Beijing Astronomical Observatory and Beijing Astrophysics Center
  of National Astronomical Observatories, Chinese Academy of Sciences.
  A20 Datun Rd., Chaoyang District, Beijing 100012;
E-mail: xue@bac.pku.edu.cn}
\altaffiltext{2}
{International School for Advanced Studies, SISSA/ISAS, via Beirut 2-4, 1-34014 
Trieste, Italy; E-mail: yhzhang@sissa.it }

%\clearpage
\begin{abstract}
We present the analysis of archival data from ROSAT and ASCA of a
serendipitous BL Lac object PKS 2316-423. Because of its featureless
non-thermal radio/optical continuum, PKS 2316-423 has been called
as a BL Lac candidate in the literature. PKS 2316-423 was evidently 
variable over the multiple X-ray observations, in particular, a variable 
high-energy tail of the synchrotron radiation is revealed.
The X-ray spectral analysis provides further evidence of the synchrotron 
nature of its broad-band spectrum: a steep and downward curving 
spectrum between 0.1--10 keV, typical of high-energy peaked BL Lacs (HBL). 
The spectral energy distribution (SED) through radio-to-X-ray yields 
the synchrotron radiation peak at frequency $\nu_p=7.3\times10^{15}$ Hz, 
with integrated luminosity of $L_{\rm syn}=2.1\times 10^{44}\rm\ ergs\ s^{-1}$. 
The averaged SED properties of PKS 2316-423 are very similar to
those ``intermediate'' BL Lac objects (IBL) found recently in several 
deep surveys, such as Deep X-ray Radio Blazar,  Radio-Emitting X-ray,
and ROSAT-Green Bank surveys.
We suggest that PKS 2316-423 is an IBL though it also shows some 
general features of a HBL. Actually, this double attribute of PKS 2316-423 
provides a good test of the prediction that an IBL object can show 
either synchrotron or inverse-Compton characteristics in different 
variability states. 

\end{abstract}

{\bf keywords:} BL Lac objects: individual (PKS 2316-423)
-- X-rays: galaxies

\clearpage
\section{Introduction}
BL Lac objects -- a subclass of active galactic 
nuclei (AGN) -- show variable non-thermal emissions from radio to
UV/X-rays, and even to $\gamma$-rays over different timescales (e.g., Urry
and Padovani 1995; Kollgaard 1994), 
which are commonly attributed to be synchrotron and inverse Compton
radiation from plasma in a relativistic 
jet oriented at a small angle with our line of sight. As such, they
represent a fortuitous natural laboratory to study the physics of jets,  
and ultimately the mechanisms of energy extraction from the central 
black holes, a fundamental goal of extragalactic astrophysics.

Earlier studies of BL Lac objects have shown that the systematic 
differences between radio and X-ray selected BL Lac objects (RBLs vs 
XBLs) are consistent with orientation differences (Kollgaard et al. 1996;
Ghisellini et al. 1993). Meanwhile, BL Lac objects 
have been reclassified in a more physical way as ``low energy'' 
and ''high energy'' peaked BL Lac objects (LBLs vs HBLs) based on the peak 
frequency of synchrotron radiation (e.g. Giommi and Padovani 1994),
rather than just obervational selection. In 
general, RBLs and XBLs tend to be LBLs and HBLs, respectively.
However, recent evidence has shown that some of the differences between
HBLs and LBLs
cannot be accounted for by differences in orientation alone (Sambruna,
Maraschi \& Urry 1996). The alternate explanation, that these classes
are merely opposite extremes of the source's spectral energy distribution 
(SED), also fails to explain some HBL-LBL differences (Kollgaard et al. 
1996; Stocke 1996). 
Instead the modern thinking is that the HBL-LBL
dichotomy represents two extremes of a continuum in either luminosity and
viewing angle (Georganopoulos \& Marscher 1998) or luminosity and peak
frequency (Fossati et al. 1998). 
     
Recent studies from deeper and larger X-ray surveys have indeed shown 
that BL Lac objects tend to exhibit more continuous distributions of 
properties (Sambruna et al. 1996; Scarpa and Falomo 1997; Perlman et al. 1998; 
Laurent-Muehleisen et al. 1999) rather than disparate ones. 
This has resulted in an important role for those objects with intermediate
SEDs between HBL and LBL, namely intermediate BL Lac objects (IBLs),
in revealing BL Lac mysteries. 

In this paper, we present the X-ray spectral analysis (ROSAT and ASCA  
archival data) and the SED of a serendipitous BL Lac object, PKS 2316-423. It 
is a southern radio source at $z=0.0549$, and was formerly classified
as a BL Lac candidate on the base of its featureless
non-thermal radio/optical continuum (Crawford \& Fabian 1994; Padovani \&
Giommi 1996). We noticed this object as it has been the brightest contaminating
source to the nearby narrow-line X-ray galaxy, NGC 7582 (Xue et al. 1998;
Schachter et al. 1998) in most of its historical X-ray records. 
 
The ROSAT(PSPC) and ASCA satellites observed this object as a serendipitous 
source in April 1993 and November 1994 respectively. These observations,
though non-simultaneous,  
extend our knowledge of the source's SED properties to the 
X-ray domain (0.1--10 keV), but also provide a good opportunity for X-ray 
spectroscopic studies, which turn out to 
be very important for its classification.

In section 2 we describe the observations and data reduction, and then 
present the spectral analysis in section 3. We construct the source
SED in section 4 on the base of our newly X-ray flux measurements as well as
the published photometry data. The results are summarized and discussed
in section 5. Throughout this paper, H$_0=50$\ \vel $\rm Mpc^{-1}$ and $q_0=0.5$
are assumed. 
All errors reported below are quoted at the 90\% confidence level 
for one interesting parameter (i.e. \delchi$=2.7$). 

\section{Observations and data reduction}

To investigate the X-ray spectroscopic properties of PKS 2316-423,
we collected the relevant data available in the NASA/GSFC data 
archive. This resulted in one 
ASCA, one ROSAT PSPC and two HRI pointed observations 
(see Table 1). All data were reduced using {\it XSELECT} within
{\it FTOOLS} package version 4.2.

PKS 2316-423 is 17 arcmin away from the center of
the ROSAT(PSPC) pointed observation, which was centered on NGC 7582.  
The source was somewhat affected by scattering and vignetting 
from the PSPC superstructure. We tried to extract the source counts 
using a larger chip region (with a radius as large as 3.2 arcmin) to 
account for former effect. The latter, which was estimated a
$\sim$9 \% effect at 1 keV, was corrected using a {\it FTOOLS}'s task,
{\it pcarf}. 
The background was estimated from an annular source-free 
region. The source spectrum
was also corrected for deadtime, and finally regrouped to at 
least 20 counts per channel. No significant variations in the source
count rate were detected over the $\sim$ 7200-s ROSAT observation
spanning a little more than one day.

Both the PKS 2316-423 and NGC 7582 were clearly detected in the
ASCA observation with the two GIS and SIS instruments. The former 
instruments were operated in normal PH mode; the later were
operated in 2 CCD mode (along the diagonal) 
and carefully oriented to place the two objects near the axis. 

The ASCA data were selected
from the intervals of high and medium telemetry rates for both GIS and SIS. 
The SIS data were screened using the following criteria: a) The data was 
not taken in the region of the South Atlantic Anomaly, b) the angle between 
the field of view and the edge of the bright and dark earth exceeded 
25$^{\circ}$ and 5$^{\circ}$, respectively,
and c) the cutoff rigidity was greater than 4 GeV\ $c^{-1}$. After these
selections, we also deleted data if d) there were any spurious events
or the dark frame error was abnormal. For GIS data, no bright earth
angle and cutoff rigidity criteria were applied. More details concerning
the performance and instrumentation of ASCA have been reported in separate
papers (ASCA satellite: Tanaka et al. 1994; SIS: Burke et al. 1991;
GIS: Ohashi et al. 1996).

Source counts of ASCA observation were extracted from circulars regions
of radius $\sim$ 3 arcmin and $\sim$ 5 arcmin for the SIS and GIS
respectively. The background was extracted from the source
free regions on the same detectors. 
No evident variations in the source
count rate were detected either over the $\sim$19,000-s ASCA observation 
spanning about half a day.

\section{Spectral analysis}

Since no evident variations in the source count rate were detected
during either observations, the time-averaged spectra from both satellites
were used for spectral analysis.
The spectra of both ROSAT PSPC and ASCA were grouped so that each energy 
channel contains at least 20 counts allowing chi-squared minimization 
techniques. Spectral analysis have been performed
using XSPEC(V10) program.\footnote{ 
\it http://legacy.gsfc.nasa.gov/docs/xanadu/xspec/u\_manual.html.}

\subsection{The ROSAT Data}

A simple power-law fit to the ROSAT PSPC data in the range of 0.1--2 keV
gives photon index $\Gamma=2.0\pm0.2$ and absorption column density of
$(1.41\pm0.5)\times10^{20}$ \pcmsq\ (Figure 1), which is consistent with the 
Galactic value
N$_{\rm Hgal}=2.0\times10^{20}\ \rm cm^{-2}$ (Stark et al.\ 1992). 
This model describes the data reasonably well (Table 2). 
The fluxes corrected for the Galactic absorption are 
$4.20^{+0.25}_{-0.28}\times10^{-12}$ \fxu\
in the 0.1--2.4 keV band and $1.83^{+0.11}_{-0.12}\times10^{-12}$ 
\ \fxu\ in the 0.5--2 keV
band, estimated from the best-fit power-law model. The inferred intrinsic
luminosity is $5.7\times10^{43}\rm\ ergs\ s^{-1}$ in the 0.1--2.4 keV band, 
which is similar to that
of other non-quasar AGN. The source was
observed twice with ROSAT HRI in June 1992 and May 1993 respectively,
the obtained count rates correspond to fluxes of 
$(5.9\pm0.3)\times10^{-12}$\ \fxu\
and $(6.1\pm0.3)\times10^{-12}$\ \fxu\ in the observed energy range 0.1-
2.4 keV, assuming a power-low spectrum identical to that of PSPC. 
%In contrast to the consistency of the source brightness recorded by the 
%HRI in an one year interval, 
The PSPC observation was made between the two HRI observations (Table 1),
and recorded a relatively lower flux level.
These factors suggest the source is variable and thus there might be a
non-thermal origin for the X-ray flux.

\subsection{The ASCA Data}
The spectra of the four ASCA instruments were fitted simultaneously in 
the range of 0.5--10 keV for SIS(0/1) and 0.7--10 keV for GIS(2/3). A simple
power-law model gives an acceptable fit to the whole dataset (Table 2).
However, comparing the results with the ROSAT data, the ASCA spectra suggest
that the source spectral slope with phonton index of $\Gamma=2.4\pm0.2$ in 
the 0.5--10 keV broad-band is significantly steeper than that of ROSAT PSPC 
spectrum in the 0.1--2 keV band. In addition, the ASCA data requires an 
absorption column density of $8.8^{+5.9}_{-5.3}\times10^{20}$\ \pcmsq\ which 
is significantly in excess of the Galactic value.    
This is clear evidence of spectral variability,
since the two observations were made 1.5 years apart. 
There are two possibe explanations for this result: either a variable
absorbing column (which, as described below, we believe to be unlikely), or
a spectral flattening at soft energies due to a convex continuum.
However, the first case seems unlikely, since that
very few \bl show evidence of significant cold absorbing gas in excess
of line-of-sight Galactic column density (Perlman et al. 1996b; Urry et al.
1996). 
The recent detections of several \bl by the Extreme Ultraviolet Explorer 
(EUVE) (Marshall, Fruscione, \&
Carone 1995; Fruscione 1996) is further evidence that these objects
do not have significant intrinsic absoption.

For the latter case, the ``excess'' absorption seen in the ASCA spectrum 
might be an artifact falsely introduced in the fitting process. 
To test this idea, we next fitted the ASCA data using a broken power-law with
a bound absorption at the Galactic value. Fitting with this model is notably 
improved (\delchi $=6.8$ for two additional parameters, $P_F>95\%$) 
compared to the fit of the single power-law model with the same bound 
absorption (Figure 2). 
The final fit yields two powerlaw components with a break-point at 
$\sim 2.1^{+0.9}_{-0.7}$ keV. The lower-energy component is flatter 
with a photon index of $\Gamma=2.0^{+0.4}_{-0.2}$; the higher-energy 
component is steeper with $\Gamma=2.6^{+0.3}_{-0.3}$. 
Thawing the absorption parameter in the broken power-law model produces
little change of chi-squared value, and the resulted absorption column 
density is still consistent with the Galactic value (Table 3).

It should be noted that although the above two models cannot be
statistically discerned, consider that the spectral-flattening
effect is by far the more likely physical explanation of the 
'excess' absorption seen in the ASCA spectra, hereafter, we 
would refer the broken power-law model as the best fit to 
the ASCA data.

Figure 3 shows confidence plots of the powerlaw spectral components
versus the corresponding absorption column density for both the ASCA and
ROSAT data.  
The ASCA observation described by the best-fitted model showed 
that the source brightness decreased by 33\% in the 0.1--2.4 keV 
range compared with ROSAT PSPC data.  

   In summary, the ASCA observation of PKS 2316-423 in November 1994 is
best-fitted by a broken power-law model, in which a 
relatively flatter component at lower energies below 
$\sim 2.1$ keV is very consistent with the shape of the ROSAT/PSPC spectrum 
observed in May 1993. This indicates that the broad-band X-ray spectral
slope remained constant at lower-energies, even though the source
brightness evidently decreased; however, the steeper component above 
$\sim 2.1$ keV does suggest the X-ray spectrum became softer
at higher energies with the decreasing of the source brightness. As 
shown in the next section, the ROSAT and ASCA observations actually revealed
a variable high-energy tail of the source synchrotron radiation.

\section{Spectral Energy Distribution}

The X-ray spectral analysis of PKS 2316-423 in the last section has shown its 
HBL-type properties. Moreover, even though non-simultaneous, the 
composite 
SED from the literature could present more insights on this object 
through comparison with the properties of the most recent complete BL Lac 
object samples. 
We plot in Figure 4 non-simultaneous radio, optical, UV and X-ray data,
from both the space and ground-based observations,
assuming an ``average'' SED of the source. It is clear that the composite 
SED from radio to X-ray is likely from only one radiation component 
(i.e., synchrotron emission). Its optical and ultraviolet 
radiation appear to be a continuation of the radio synchrotron spectrum;
the X-ray data are likely from a common emission origin as the lower 
energy parts and represents a high energy tail of the synchrotron spectrum. 

First, we derive some spectral parameters from the SED of PKS 2316-423. 
Following the general definition, we get the K-corrected two 
point spectral indices: radio-optical spectral index $\alpha_{ro}=0.56$, 
optical-X-ray one $\alpha_{ox}=1.18/1.26$ and radio-X-ray one 
$\alpha_{rx}=0.78/0.82$ (note that hereafter the two values of one parameter 
is due to the different X-ray fluxes discussed in the last section).   
Because of the spectral variability shown evidently in the hard X-ray band (
$>2$ keV), the composite X-ray/optical spectral, $\alpha_{xox} 
=\alpha_{ox}-\alpha_x$ also varied, with a value of $0.18/0.26$ for
$\alpha_x=1.0$\footnote{Note that the parameter used here
refers to the energy spectral index, which is related to the photon 
spectral index with $\alpha_x=\Gamma -1$.}, and 
$\alpha_{xox} = -0.42/-0.34$ for $\alpha_x=1.6$, respectively. Another 
important spectral parameter is the 
X-ray to radio logarithmic flux density ratio, and we get log$S_X/S_r 
\sim -6.36/-6.29$. 

Furthermore, it is clearly seen from Figure 4 that the radio to X-ray SED 
of PKS 2316-423 likely peaks  
in the EUV/soft-X-ray band. Following Sambruna et al. (1996), we 
performed a parabolic fit to the radio through optical/UV/X-ray
SED of the source to determine the value of the peak
frequency. We obtain $\nu_p=7.3\times10^{15}$ Hz. 
The same fit yields an estimate of the luminosity at the peak
frequency $L_p=1.5\times10^{43}\rm\ ergs\ s^{-1}$
and of the integrated radio-to-X-ray synchrotron
luminosity of $L_{\rm syn}=2.1\times 10^{44}\ \rm ergs\ s^{-1}$. 
In the next section we will discuss the general SED properties of BL Lac 
objects and the attribute of PKS 2316-423.

For comparison, we plotted in Figure 4 
the EGRET sensitivity threshold as an upper limit to the GeV flux 
(marked by an arrow), since the source was never detected at $\gamma$-rays. 
It is shown that the source is dominated by a synchrotron process in
the a broad energy range below 10 keV; above which, we know nothing
about the inverse-Compton emission which would undoubtedly dominate the
source emission.

\section{Discussions}

PKS 2316-423 was first serendipitously detected as a marginally
extended (1.5 arcmin) X-ray source in the Einstein Medium-Sensitivity
Survey (EMSS; Gioia et al. 1984; Stocke et al. 1991). The ROSAT HRI
observation made by Crawford and Fabian (1994) in 1992 June confirmed 
the result of Einstein and further revealed a second weak X-ray source 
10 arcsec (15 kpc) south of the galaxy which may be related to the 
central source. They suggested  that PKS 2316-423 is a BL Lac object 
based on its probably non-thermal radio/optical/X-ray radiation, and 
has an ``off-sight'' jet component. 
If this is verified, then it will provide an interesting test case
for both the host galaxies of BL Lac objects and their environment.
Unfortunately, the later ROSAT HRI observation made in 1993 May
did not go deep enough compared to the former observation, thus 
the misaligned scenario of this BL Lac object can hardly be 
further confirmed or constrained. 

The multiple X-ray observations show that PKS 2316-423 was variable  
on timescales of weeks to years. The flux in the 0.1--2.4
keV band increased by $\sim$50 \% from ROSAT PSPC to HRI observations
in a 13 day interval in May 1993, and dropped  
$\sim$73 \% 1.5 years later when observed with ASCA in November 1994. 
These factors suggest the source is variable and thus consistent with
the expected characteristics of a high-energy tail from a dominant 
synchrotron emission.
The source, however, did not show any short term variability during 
both ROSAT PSPC and ASCA observations spanned over $\sim$ 1 day. More 
X-ray observations will be necessary. 

X-ray spectral analysis provides more insights into the nature of
the source's emission. A single power-law model fit to the ROSAT PSPC
and ASCA data shows significant discrepancies. However, the two datasets
can be reconciled if the ASCA spectrum is fit with
a broken power-law model. This model indicates that, below an 
energy-break point, $\sim 2.1$ keV, there is a relatively flatter component 
($\Gamma=2.0^{+0.4}_{-0.2}$) to be consistent with the PSPC spectrum; 
and a steeper component with $\Gamma=2.6^{+0.3}_{-0.3}$ above 
the energy-break point, this is one of the general X-ray spectral properties of 
a HBL (Sambruna et al. 1996). This kind of intrinsically downward 
curved X-ray spectrum can be easily interpreted as due to   
synchrotron losses from a relativistic plasma. The high-energy peaked SED of 
the source indicates that PKS 2316-423 is a HBL-type BL Lac 
object, with the X-ray spectrum being 
the high energy tail of synchrotron emission over a wide wavelength range, 
with correlated flux and spectral variability.  
With the decrease in source brightness from ROSAT PSPC
to ASCA observations, the X-ray spectrum in 0.1--10 keV band became softer 
in the harder X-ray band.

The recent studies have greatly modified our view of BL Lac objects. DXRBS 
(Perlman et al. 1998), REX (Caccianiga et al. 1999), and RGB 
(Laurent-Muehleisen et al. 1999) surveys have shown that BL Lac objects 
exhibit a continuous distribution of properties rather than two 
distinct classes: HBLs vs 
LBLs (or XBLs vs RBLs). These findings show that the distribution of the 
SED parameters of BL Lac objects peaks where the empty region 
between the two extreme subclasses were previously seen. Therefore, most BL 
Lac objects should exhibit intermediate SED properties. The previously 
observed bimodal distribution is primarily due to observational selection
effects (e.g. Laurent-Muehleisen et al. 1999).

Here we compare the SED properties of PKS 2316-423 with the general 
properties of BL Lacs found recently. When we put PKS 2316-423 
($\alpha_{ro}=0.56$ and $\alpha_{ox}=1.18/1.26$) on 
the $\alpha_{ro}$ vs $\alpha_{ox}$ color-color diagram, we find that it 
is a somewhat intermediate BL Lac object. As we know, 
$\alpha_{XOX}$ can more precisely measure spectral changes from optical to 
soft X-ray bands.
If $\alpha_{XOX}\lta 0$, then the X-rays lie along a powerlaw or
steepening synchrotron continuum. A positive value of $\alpha_{XOX}$
represents a concave spectrum and is likely caused by a hard
inverse-Compton component.
The values of $\alpha_{XOX}$ for PKS 2316-423 
depend on the X-ray spectral slopes in different variability
states, being 0.18/0.26 
and -0.42/-0.34 for $\alpha_x=1.0$, in the high state of the ROSAT/PSPC 
observation, and 1.6, in the low
state of the ASCA observation respectively. We find these 
values should just locate in the intermediate range of the $\alpha_{XOX}$ 
distribution of recent BL Lac samples (e.g., Perlman et al. 1998; 
Caccianiga et al. 1999; Laurent-Muehleisen et al. 1999).
Interestingly, the value of $\alpha_{XOX}$ at different epoch was in 
opposite symbol, this might indicate that, in the different 
variability states of PKS 2316-423, the spectral change from optical 
to X-ray could show either synchrotron or inverse-Compton characteristics, 
just as would be expected for an IBL.

Previous studies show a clear bimodality in the ratio of the X-ray 
to radio flux densities of HBLs and LBLs at log$S_X/Sr \sim -5.5$ (e.g. 
Padovani and Giommi 1995; Perlman et al. 1996a; Brinkmann et al. 1996). 
The value of log$S_X/Sr$ 
for PKS 2316-423 is -6.36/-6.29, smaller than this sharp division value, 
seems in contradiction with its non-LBL attribute; 
however, the value is consistent with the distribution
of the flux ratios of the 
intermediate BL Lacs found in DXRBS and RGB samples, which show no such 
dichotomy between HBL- and LBL-like SEDs (Perlman et al. 1998; 
Laurent-Muehleisen et al. 1999). 

The importance of the frequency at which the synchrotron radiation 
peaks is that it provides a powerful diagnostic for the physical 
condition of the emitting region. Recent studies showed that among BL 
Lacs the synchrotron peak frequencies are inversely correlated with their 
luminosities (Fossati et al. 1998). Due to its low peak luminosity,
PKS 2316-423 should be located near the bottom right end of Figure 7c
of Fossati et al. (1998), with a peak frequency of around $\sim 10^{18}$
Hz. However the parabolic fit to the SED of PKS 2316-423 just gives $\nu_p
\sim 10^{16}$ Hz. Consider the given 
ROSAT photon index of 2.0 (see section 3), which equates to a flat spectrum 
in $\nu F_{\nu}$ vs $\nu$ space within the soft X-ray range, suggesting  
that the peak frequency of PKS 2316-423 might be underestimated. This
could be caused by imposing that the X-ray point smoothly
connects to the lower energy data, as would happen in a parabolic
fit. If a cubic fit is adopted, the synchrotron peak of PKS 2316-423 
could be shifted toward a value of $\sim 10^{17}$ Hz, which is 
basically follows the general trend of inverse correlation of
the synchrotron peak frequency versus the luminosity 
among blazars (Fossati et al. 1998). In this sense,
however, the source also behaves like a typical HBL. 

In summary, the averaged SED properties of PKS 2316-423 indicates  
that it is an IBL, like those many objects found in DXRBS (Perlman et al. 1998), 
REX (Caccianiga et al. 1999) and RGB surveys (Laurent-Muehleisen et al. 1999). 
However, the source also shows some typical X-ray spectral properties of a HBL.
This in fact supports a people's general idea -- that in X-rays, 
an intermediate BL Lac object can show either synchrotron or
inverse-Compton characteristics in different variability states.
Though the most compelling test of the idea awaits a large sample 
of such studies.

\acknowledgments
%\vskip 0.5cm
We would like to thank the anonymous referee for his/her 
insightful comments that led to a number of significant
improvements in this paper. This research has 
made use of the NASA/IPAC Extra-galactic Database (NED) which is 
operated by the Jet Propulsion Laboratory, California Institute 
of Technology, under contract with the National 
Aeronautics and Space Administration. S.J.X. acknowledges the 
financial support from the Chinese Post-Doctoral Program.

\begin{deluxetable}{lccc}
\small
\tablecaption{Summary of ROSAT and ASCA data}
\tablehead{
\colhead{Instrument} & \colhead{Date} & \colhead{Exposure} &\colhead{Count rate}\\
& \colhead{} & \colhead{(ksec)} & \colhead{(count\ s$^{-1}$)} }
\startdata
ROSAT PSPC & 93/05/04  & 7.2 & $0.228\pm0.006$  \nl
ROSAT HRI & 92/06/07  & 7.5 & $0.065\pm0.003$  \nl
ROSAT HRI & 93/05/17  & 4.2 & $0.067\pm0.004$  \nl
ASCA SIS0 & 94/11/14  & 16.4 &$0.047\pm0.002$  \nl
ASCA SIS1 &  & 16.4 & $0.034\pm0.002$  \nl
ASCA GIS2 &  & 18.8 & $0.024\pm0.002$  \nl
ASCA GIS3 &  & 18.8 & $0.026\pm0.002$  \nl
\enddata
\tablenotetext{} {}
\end{deluxetable}

\begin{deluxetable}{cccccc}
\small
\tablecaption{Simple power-law fits to the ROSAT PSPC and ASCA data}
\tablehead{
\colhead{Data} & \colhead{$\Gamma$} &\colhead{$N_{\rm H}$} & \colhead{\rchi/d.o.f} &
\colhead{F$_{1\rm keV}$\tablenotemark{a}} & \colhead{F$_{0.1-2\rm keV}$\tablenotemark{a}} 
\\
& & \colhead{($10^{20}$\pcmsq)} & & \colhead{($10^{-12}$\fxu)} 
& \colhead{($10^{-12}$\fxu)}  }
\startdata
ROSAT PSPC & 2.0$^{+0.2}_{-0.2}$  & 1.41$^{+0.49}_{-0.43}$ & 1.1/17 
& $1.32^{+0.08}_{-0.09}$ & $2.63^{+0.15}_{-0.18}$ \nl 
ASCA & 2.4$^{+0.2}_{-0.2}$  & 8.78$^{+5.97}_{-5.37}$ & 0.96/133 
& $1.10^{+0.16}_{-0.19}$ & $1.35^{+0.33}_{-0.20}$ \nl
\enddata
\tablenotetext{a} {Observed fluxes (not corrected for absorption).}
\end{deluxetable}

\begin{deluxetable}{ccccccc}
\small
\tablecaption{Broken power-law fit to the ASCA data}
\tablehead{
\colhead{\nh} & \colhead{$\Gamma1$} & \colhead{$E_{break}$} & \colhead{$\Gamma2$} 
& \colhead{\rchi/d.o.f} & \colhead{F$_{0.5-2\rm keV}$\tablenotemark{a}} 
& \colhead{F$_{2-10\rm keV}$\tablenotemark{a}} \\
($10^{20}$ \pcmsq)& & \colhead{(keV)} & & & \colhead{($10^{-12}$ \fxu)} 
& \colhead{($10^{-12}$ \fxu)}  }
\startdata
2.4$^{+2.2}_{-2.0}$ & 2.0$^{+0.4}_{-0.2}$ & 2.1$^{+0.9}_{-0.7}$ & 2.6$^{+0.3}_{-0.3}$ & 0.95/131 
& $1.13^{+0.34}_{-0.16}$ & $0.92^{+0.28}_{-0.14}$ \nl
\enddata
\tablenotetext{a} {Observed fluxes (not corrected for absorption).}
\end{deluxetable}

\clearpage
\begin{figure}
\hbox{
%\hspace{0.6cm}
\centerline{\psfig{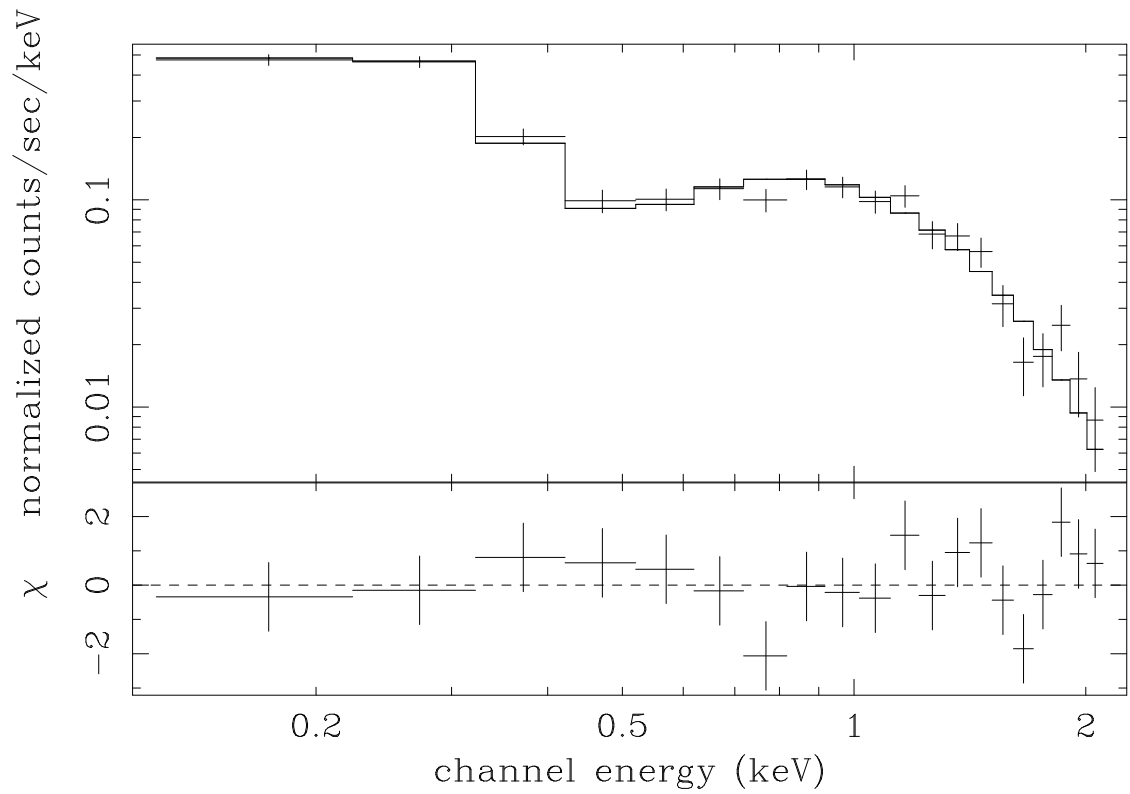}}
}
\caption{
Folded ROSAT PSPC spectrum of PKS 2316-423 in May 1993. Single power-law fit 
to the observation and residuals are shown. The model gives acceptable 
fit to the data with an absorption in good agreement with the Galactic value.}
\end{figure}

\clearpage
\begin{figure}
\hbox{
%\hspace{0.6cm}
\centerline{\psfig{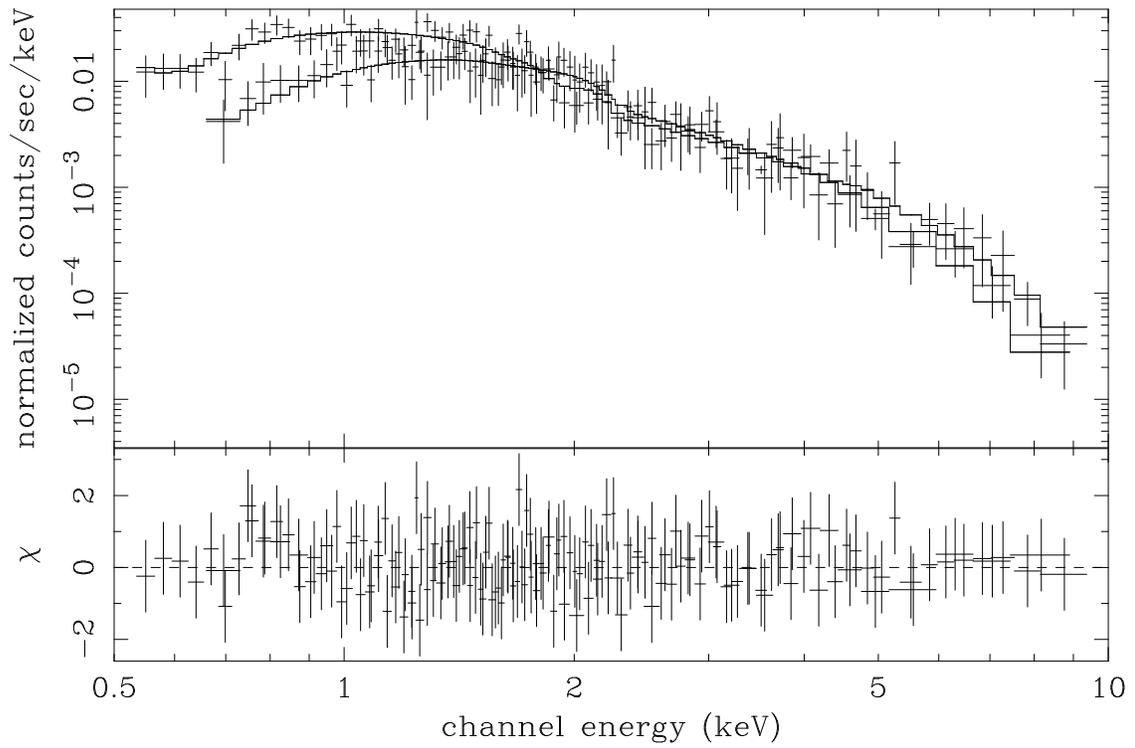}}
}
\caption{Folded ASCA spectra of PKS 2316-423 in November 1994. Data
from all the four detectors were best fitted together with
a broken power-law model (see text). Only combined
SIS and GIS spectra are shown for conciseness. 
}

\end{figure}

\clearpage
\begin{figure}
\hbox{
%\hspace{0.6cm}
\centerline{\psfig{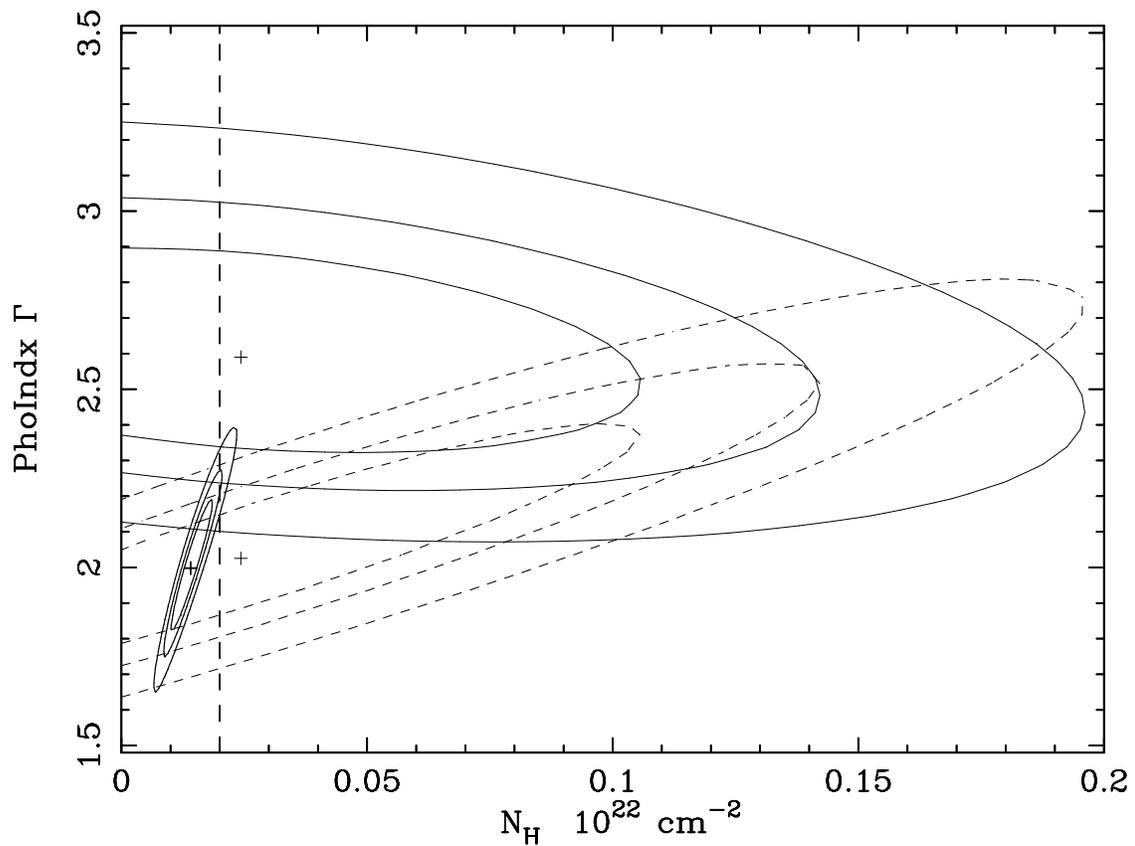}}
}
\caption{
Confidence contour levels (68\%, 90\% and 99\%) for the 
slope of power-law components vs. the absorption column density.
The thin solid and dashed lines show the steep and flat components
of the ASCA spectrum respectively. Thick solid lines show for the 
ROSAT/PSPC spectrum. The vertical dashed line shows the absorption
column density at the Galactic value.}
\end{figure}

\clearpage
\begin{figure}
\hbox{
%\hspace{0.6cm}
\centerline{\psfig{figure=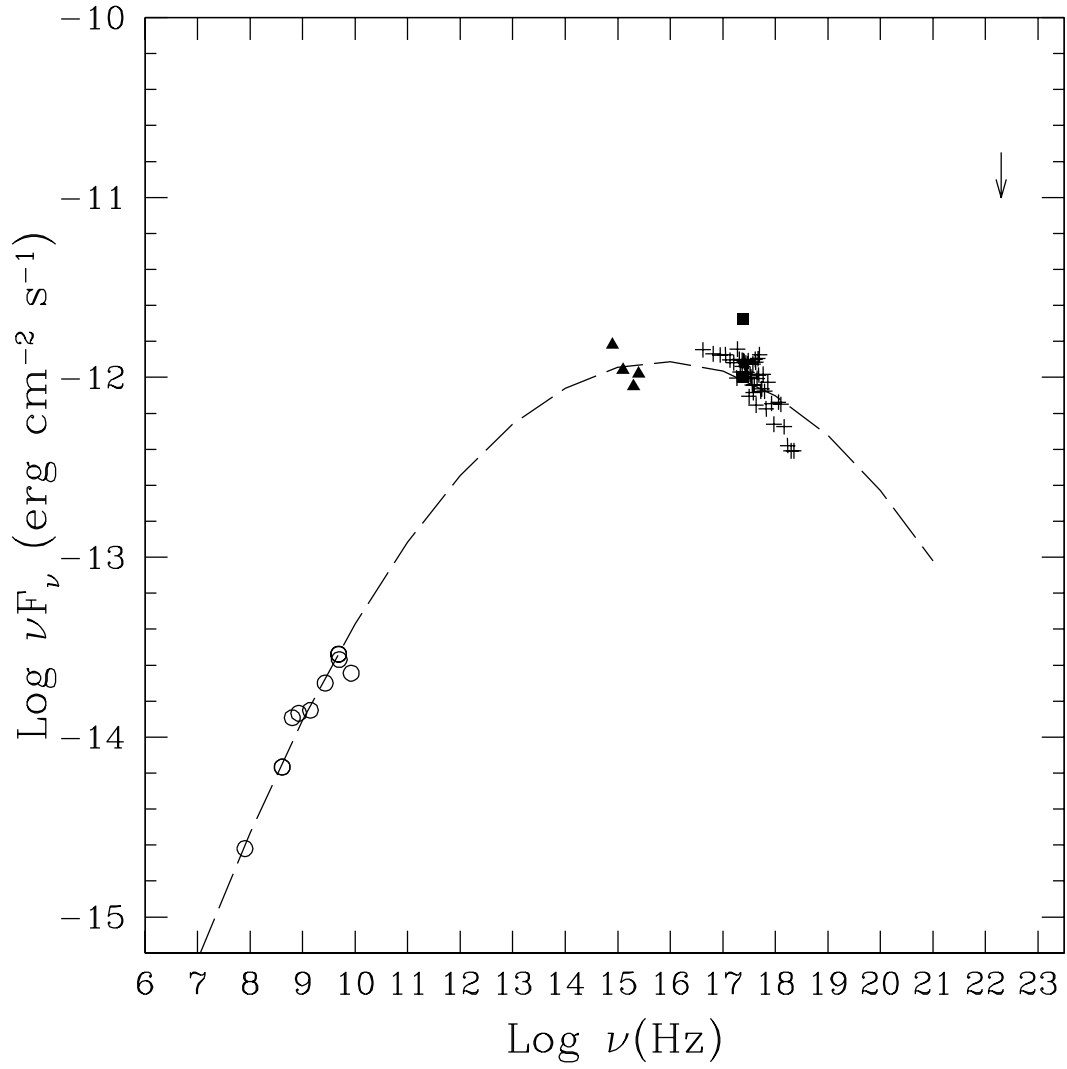,width=0.8\textwidth,angle=0}}
}
\caption{
The multifrequency SED of PKS 2316-423 and its parabolic fit. The 
X-ray points are data from this paper, plotted with solid square 
for ROSAT/HRI and cross mark for ASCA and ROSAT/PSPC.
The UV/optical points are data from Crawford and Fabian (1994) plotted 
with solid triangle. Circle symbols represent radio data
from NASA/IPAC Extra-galactic Database (NED).}
\end{figure}
\end{document}